\documentclass[%
 preprint, 
 amsmath,amssymb,
 aps, physrev,
]{revtex4-2}
\usepackage{physics}
\usepackage{tikz}
\usepackage{subcaption}
\usepackage{mathtools}
\usepackage{verbatim}
\usepackage{graphicx}
\usepackage{dcolumn}
\usepackage{bm}
\usepackage{hyperref}

\usepackage{lipsum}
\usepackage{xcolor}
\usepackage{mathrsfs}
\begin{document}


\title{\textbf{Thermal nature of the causal diamond horizon:\\A hidden property of the inertial propagator} 
}%

\author{Nada Eissa,$^{1}$ Carlos R. Ordóñez,$^{1}$ and Gustavo Valdivia-Mera$^{1}$}

\affiliation{\vspace{5mm}
$^{1}$Department of Physics, University of Houston, Houston, Texas  77204-5005, USA\vspace{5mm}}
\date{\today}\vspace{5mm}
\begin{abstract}
Inspired by the novel idea proposed by T.~Padmanabhan in \textit{Phys.\ Rev.\ D 100, 045024 (2019)}, we develop a method to uncover the hidden thermal properties of the inertial Feynman propagator in Minkowski spacetime in a causally consistent manner. This, in turn, enables a coherent interpretation based on future-directed propagation. In our approach, the Fourier transform is implemented following the convention used in the analysis of vacuum fluctuations. As a result, future-directed propagation across causal horizons can be consistently interpreted, from the perspective of an observer confined to a causally disconnected region, as the emission of scalar quanta at the past horizon and their absorption at the future horizon. Moreover, we find that the ratio between emission and absorption processes reproduces the characteristic Boltzmann factor of a thermal ensemble. We first apply this analysis to a causal diamond of length $2\alpha$, performing a detailed study of the near-horizon geometry and thereby obtaining the temperature associated with the thermal behavior of the Minkowski vacuum as perceived by an observer with finite lifetime $2\alpha$. For completeness, we also apply the method to the right Rindler wedge, recovering the well-known Unruh temperature, $T = a/(2\pi)$. Our results demonstrate that thermality can emerge directly from causal structure, independently of acceleration or gravity, with causal diamonds encoding intrinsic thermodynamic behavior in quantum field theory.
\end{abstract}

\maketitle
\newpage
\tableofcontents

\newpage
\section{Introduction}
In his insightful work, T.~Padmanabhan \cite{padmanabhan2019thermality} showed that the inertial Feynman propagator for a real scalar field in Minkowski spacetime loses its time-reversal symmetry in the presence of causal horizons. Moreover, by considering the Fourier transform of the Feynman propagator with respect to the temporal interval, he interpreted backward-in-time propagation from the future wedge $F$ to the right wedge $R$ as the emission of scalar quanta from the horizon, while forward-in-time propagation from $R$ to $F$ was interpreted as absorption. Remarkably, the squared norm of the ratio of these Fourier-transformed propagators yields a thermal Boltzmann factor that reproduces the well-known Unruh temperature, $T = a/(2\pi)$, for uniformly accelerated observers \cite{fulling1973nonuniqueness,davies1975scalar,unruh1976notes}.\\

In this article, we introduce a new method inspired by the approach of Ref.\cite{padmanabhan2019thermality}, which employs the Feynman propagator to describe physical processes across causal horizons. Our formulation departs conceptually from that framework by implementing a physically motivated and causally consistent prescription for the temporal Fourier transform, aligned with the conventions used in the analysis of vacuum fluctuations~\cite{birrell1984quantum,DeWitt:2003pm}. Within this setting, we restrict attention to future-directed propagations between timelike separated events, interpreting transitions across the causal horizon as emission or absorption processes of scalar quanta. By establishing a triad of events that links these processes, our method provides a self-consistent, causally sound framework that unifies the propagator-based description with the physical interpretation of horizon thermality. This construction represents a physically grounded extension of the propagator approach to horizon thermodynamics.

This article is organized as follows: In Sec.~II, we present the Feynman propagator in a representation that emphasizes its dependence on squared spacetime intervals and enables a useful Fourier transform in the time interval, followed by an analysis of the geometry of the causal diamond, focusing on its boundary structure in Minkowski spacetime and its near-horizon regime. In Sec.~III, we describe the method that will be used in the remainder of the section, and then apply it to both the causal diamond and Rindler cases. We conclude in Sec.~IV with a discussion of our results and their implications for the emergence of thermality from causal structure.

\section{The Feynman propagator and causal diamond geometry}
\subsection{The Feynman propagator and its frequency-space representation} \label{secii}

The Feynman propagator in $d$-dimensional Minkowski spacetime for a real massive scalar field, with metric signature $(-,+,+,\cdots)$, can be expressed as
\begin{eqnarray}
G_F(x - y) &=& i \left(\frac{1}{4\pi i}\right)^{d/2} \int_0^\infty \frac{ds}{s^{d/2}} \exp\left[\frac{i}{4s} \sigma^2(x,y) - i s m^2 - s \epsilon \right], \label{schw122}
\end{eqnarray}
where $\epsilon \to 0^+$ enforces the appropriate boundary conditions for time ordering, and $\sigma^2(x,y)$ denotes the squared Minkowski spacetime interval between the events $x^\mu$ and $y^\mu$ (for a detailed discussion on how to recover the standard form of the Feynman propagator as presented in Refs.~\cite{weinberg1995quantum,srednicki2007quantum}, see Ref.~\cite{padmanabhan2016quantum}). In addition, the dependence of the Feynman propagator $G_F$ on the temporal coordinates arises solely through the interval $\Delta t = x^0 - y^0$, which is a direct consequence of the static nature of Minkowski spacetime.\\

In what follows, it will prove advantageous to work with a representation in which the time dependence of the propagator is transferred to its conjugate frequency domain. This is particularly useful when analyzing propagation across causal horizons or when exploiting the symmetries of static spacetimes. To this end, we introduce the \emph{frequency-space propagator} by performing a Fourier transform of $G_F$ with respect to the time interval, following the standard procedure for vacuum fluctuation detection~\cite{birrell1984quantum,DeWitt:2003pm}:
\begin{equation}
\tilde{G}_F(\Omega, \mathbf{x}, \mathbf{y}) = \int_{-\infty}^\infty d(\Delta t)\, G_F(\Delta t, \mathbf{x}, \mathbf{y})\, e^{-i \Omega \Delta t}, \label{foutregf2}
\end{equation}
where $\Omega$ is the frequency conjugate to the time separation $\Delta t$.
In Minkowski spacetime, the analytic structure of $\tilde{G}_F(\Omega, \mathbf{x}, \mathbf{y})$ is characterized by simple poles at $\Omega = \pm(\omega_{\mathbf{k}} - i\epsilon)$. These singularities encode the causal and time-ordered nature of $G_F$. By restricting attention to $\Omega > 0$, we isolate the contribution of positive-frequency modes, which are naturally associated with physical excitations propagating forward in time. 

\subsection{Causal diamond geometry}
\subsubsection{Causal diamond spacetime from Rindler}
\begin{figure}[h]
 \centering
   \begin{subfigure}{0.35\textwidth}
    \includegraphics[width=\linewidth]{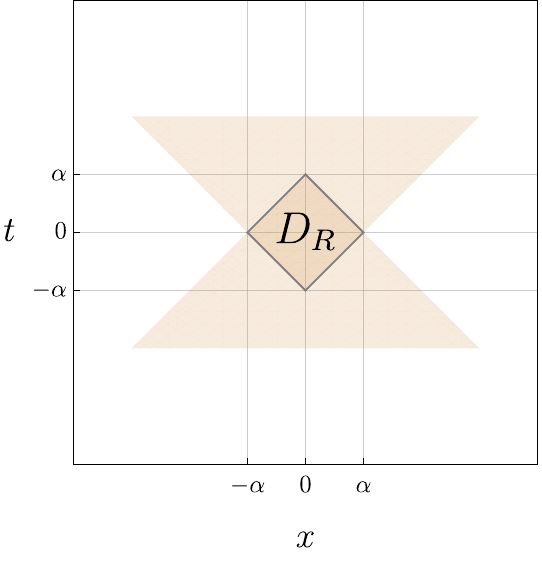}
     \caption{}\label{cfcp}
   \end{subfigure}
   \hspace{0.5cm}
   \begin{subfigure}{0.335\textwidth}
     \centering
     \includegraphics[width=\linewidth]{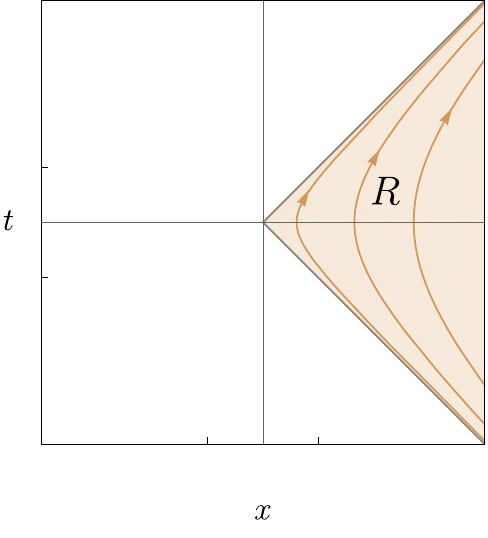}
     \caption{}\label{rrw}
   \end{subfigure}
  \caption{(a) Causal diamond $D_R$ defined as the intersection of a future and a past light cone. (b) Uniformly accelerated trajectories in the right Rindler wedge $R$.}
\end{figure}

A causal diamond of length $2\alpha$ in Minkowski spacetime is given by the region $D_R$, defined as $D_R := \{(t,x) : \abs{t} + \abs{x} \leq \alpha\}$. This region corresponds to the causal domain of an observer with a finite lifetime, bounded by the intersection of the future light cone of the birth event at $(t,x) = (-\alpha, 0)$ and the past light cone of the death event at $(t,x) = (\alpha, 0)$, as illustrated in Fig.~\ref{cfcp}. Therefore, the boundary of the causal diamond precisely contains the causal horizon associated with this observer.\\

The spacetime experienced by such a finite-lifetime observer defines the so-called \emph{Causal Diamond} (CD) spacetime. It can be geometrically characterized from the subset of Minkowski spacetime points restricted to the region $D_R$, denoted as $x^\mu = (t_{d_R}, x_{d_R})$. These coordinates are obtained by applying the conformal transformation introduced in Ref.~\cite{camblong2024entanglement} to the Minkowski coordinates restricted to the right Rindler wedge $R$ (see Fig.~\ref{rrw}), which are given by $\tilde{x}^\mu = (t_r, x_r)$. Accordingly, the conformal map is defined as
\begin{equation}
\tilde{x}^\mu \mapsto x^\mu = \qty[T \circ K \circ D]\, \tilde{x}^\mu.
\label{cm1}
\end{equation}

The transformation introduced in Eq.~\eqref{cm1} is given by the following finite conformal group transformations (see Ref. \cite{francesco2012conformal}):
\begin{align}
\text{Dilation:} \qquad & D(\lambda)\tilde{x}^\mu = \lambda\, \tilde{x}^\mu,\label{dktlbcdd}\\
\text{Translation:} \qquad & T(c)\tilde{x}^\mu = \tilde{x}^\mu + c^\mu,\label{dktlbctcc}\\
\text{Special Conformal Transformation (SCT):} \qquad & K(b)\tilde{x}^\mu = \frac{\tilde{x}^\mu - b^\mu (\tilde{x} \cdot \tilde{x})}{1 - 2(b \cdot \tilde{x}) + (b \cdot b)(\tilde{x} \cdot \tilde{x})}.
\label{dktlbc}
\end{align}
For later use, the parameters of the SCT and the translation are fixed as
\begin{equation}
b^\mu \equiv \left(0, -\frac{1}{2\alpha}\right)\quad, \quad 
c^\mu \equiv (0, -\alpha).
\label{pm1}
\end{equation}

Now, consider the coordinates $(t_r, x_r)$ defined through the following suitably rescaled transformation between Minkowski and Rindler coordinates $(\eta, \rho)$ (for a more detailed discussion, see Ref.~\cite{camblong2024entanglement}):
\begin{equation}
\frac{t_r}{\tilde{\alpha}} = \qty(\frac{2}{\alpha})\rho \sinh\left(\frac{2\eta}{\alpha}\right)\quad,\quad \frac{x_r}{\tilde{\alpha}} = \qty(\frac{2}{\alpha})\rho\cosh\left(\frac{2\eta}{\alpha}\right),\label{trrhoeta1}
\end{equation}
where $\tilde{\alpha} = 2\alpha/\lambda$. Then, from Eqs.~\eqref{cm1}–\eqref{pm1}, we obtain the coordinate transformation between the Minkowski points restricted to the region $D_R$ and the CD spacetime coordinates $(\eta, \rho)$:
\begin{equation}
    t_{d_r} = \frac{4 \alpha^2 \rho \sinh \left(\frac{2 \eta }{\alpha} \right)}{\alpha^2 + 4\alpha\rho \cosh \left(\frac{2 \eta }{\alpha} \right) + 4\rho^2}\quad,\quad
    x_{d_r} = \frac{4\alpha\rho^2 - \alpha^3}{\alpha^2 + 4\alpha\rho \cosh \left(\frac{2 \eta }{\alpha} \right) + 4\rho^2},
\label{tdrxdretarho}
\end{equation}

It is worth noting that the dilation parameter $\lambda$ does not appear explicitly in Eq.~\eqref{tdrxdretarho}, as it has been absorbed into the rescaled coordinate transformation between Minkowski and Rindler spacetimes introduced in Eq.~\eqref{trrhoeta1}.\\

The line element of the causal diamond (CD) spacetime takes the form
\begin{equation}
ds^2 = \Lambda^2(\eta,\rho) \left[ -\frac{4\rho^2}{\alpha^2} \, d\eta^2 + d\rho^2 \right], \label{cdstm1}
\end{equation}
where $\eta \in (-\infty, \infty)$ and $\rho > 0$, with the conformal factor given by
\begin{equation}
\Lambda^2(\eta,\rho) = \frac{16 \alpha^4}{\left( \alpha^2 + 4 \alpha \rho \cosh\left( \frac{2\eta}{\alpha} \right) + 4 \rho^2 \right)^2}.
\label{cdstm1cf}
\end{equation}

\subsubsection{Complementary regions: wedges and diamonds}
\begin{figure}[h]
 \centering
   \begin{subfigure}{0.33\textwidth}
    \includegraphics[width=\linewidth]{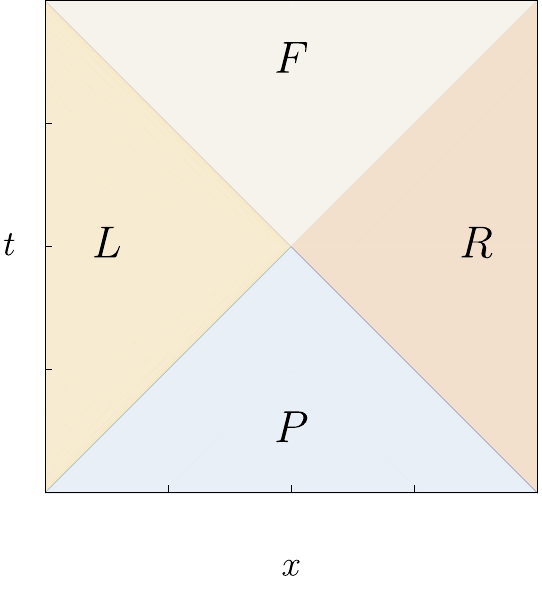}
     \caption{}\label{r4fig}
   \end{subfigure}
   \hspace{0.5cm}
   \begin{subfigure}{0.35\textwidth}
     \centering     \includegraphics[width=\linewidth]{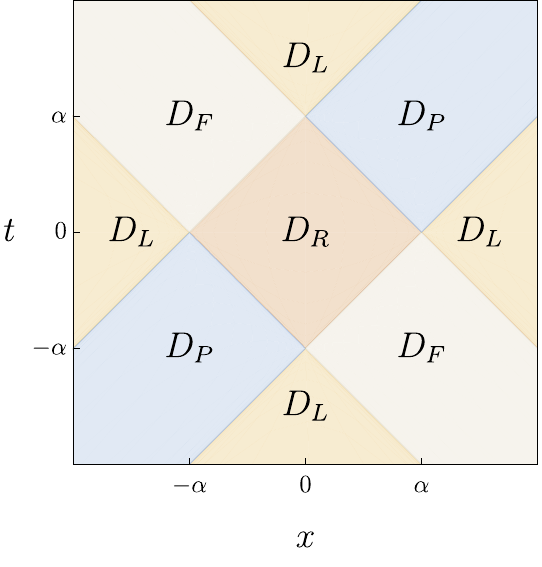}
     \caption{}\label{r4d4}
   \end{subfigure}
  \caption{(a) Rindler wedges, $R$, $L$, $P$, and $F$, in Minkowski spacetime. (b) Diamond regions $D_R$, $D_L$, $D_P$, and $D_F$ in Minkowski spacetime.}
\end{figure}

Minkowski spacetime contains not only the right Rindler wedge $R$, but also the left $L$, past $P$, and future $F$ wedges, as shown in Fig.~\ref{r4fig}. These additional regions, although regarded as unphysical due to their unconventional causal structures, are nevertheless geometrically meaningful in the context of an extended conformal mapping. Accordingly, just as we characterized the region $D_R$ from the wedge $R$ via the conformal map in Eq.~\eqref{cm1}, we now construct the mappings from the wedges $L$, $P$, and $F$ to their corresponding diamond regions $D_L$, $D_P$, and $D_F$, respectively, as color-coded in Fig.~\ref{r4d4}.

To do so, we begin with the suitably rescaled transformations, analogous to those defined in Ref.~\cite{camblong2024entanglement}, between the Minkowski and Rindler coordinates associated with the left, past, and future Rindler wedges:
\begin{eqnarray}
    \frac{t_l}{\tilde{\alpha}} = -\frac{2\rho}{\alpha} \sinh\left(\frac{2\eta}{\alpha}\right) &\quad ,\quad& \frac{x_l}{\tilde{\alpha}} = -\frac{2\rho}{\alpha} \cosh\left(\frac{2\eta}{\alpha}\right),\label{trrhoeta2}
\end{eqnarray}
\begin{eqnarray}
    \frac{t_f}{\tilde{\alpha}} = \frac{2\rho}{\alpha} \cosh\left(\frac{2\eta}{\alpha}\right) &\quad ,\quad& \frac{x_f}{\tilde{\alpha}} = \frac{2\rho}{\alpha} \sinh\left(\frac{2\eta}{\alpha}\right),\label{trrhoeta3}
\end{eqnarray}
\begin{eqnarray}
    \frac{t_p}{\tilde{\alpha}} = -\frac{2\rho}{\alpha} \cosh\left(\frac{2\eta}{\alpha}\right) &\quad ,\quad& \frac{x_p}{\tilde{\alpha}} = -\frac{2\rho}{\alpha} \sinh\left(\frac{2\eta}{\alpha}\right).\label{trrhoeta4}
\end{eqnarray}

Applying the conformal map~\eqref{cm1}, we obtain the Minkowski coordinates restricted to the regions $D_L$, $D_P$, and $D_F$, denoted by $(t,x) = (t_{d_i}, x_{d_i})$ with $i \in \{l, p, f\}$, expressed in terms of the causal diamond coordinates $(\eta,\rho)$:
\begin{eqnarray}
t_{d_l} = \frac{-4\alpha^2\rho\sinh(\frac{2\eta}{\alpha})}{\alpha^2 - 4\alpha\rho\cosh(\frac{2\eta}{\alpha}) + 4\rho^2}\quad,\quad
x_{d_l} = \frac{4\alpha\rho^2 - \alpha^3}{\alpha^2 - 4\alpha\rho\cosh(\frac{2\eta}{\alpha}) + 4\rho^2},\label{minktdl}
\end{eqnarray}
\begin{eqnarray}
t_{d_f} = \frac{4\alpha^2\rho\cosh(\frac{2\eta}{\alpha})}{\alpha^2 + 4\alpha\rho\sinh(\frac{2\eta}{\alpha}) - 4\rho^2}\quad,\quad
x_{d_f} = \frac{-4\alpha\rho^2 - \alpha^3}{\alpha^2 + 4\alpha\rho\sinh(\frac{2\eta}{\alpha}) - 4\rho^2},\label{tdrxdretarhodff1}
\end{eqnarray}
\begin{eqnarray}
t_{d_p} = \frac{-4\alpha^2\rho\cosh(\frac{2\eta}{\alpha})}{\alpha^2 - 4\alpha\rho\sinh(\frac{2\eta}{\alpha}) - 4\rho^2}\quad,\quad
x_{d_p} = \frac{-4\alpha\rho^2 - \alpha^3}{\alpha^2 - 4\alpha\rho\sinh(\frac{2\eta}{\alpha}) - 4\rho^2}.\label{tdrxdretarhodff}
\end{eqnarray}

\subsubsection{Boundary of the causal diamond}\label{bbrdr}
\begin{figure}[h]
 \centering
   \begin{subfigure}{0.35\textwidth}
    \includegraphics[width=\linewidth]{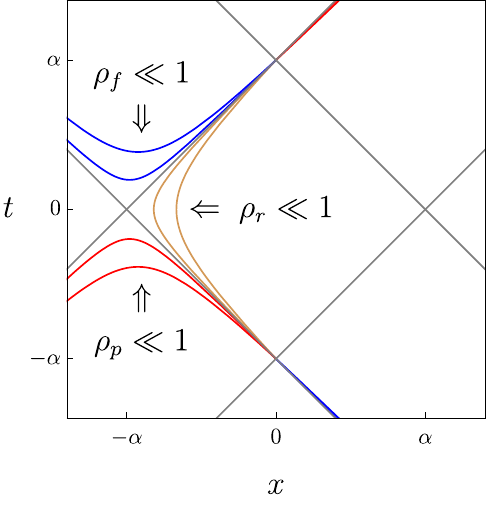}
     \caption{}\label{nhapfig}
   \end{subfigure}
   \hspace{0.5cm}
   \begin{subfigure}{0.35\textwidth}
     \centering
     \includegraphics[width=\linewidth]{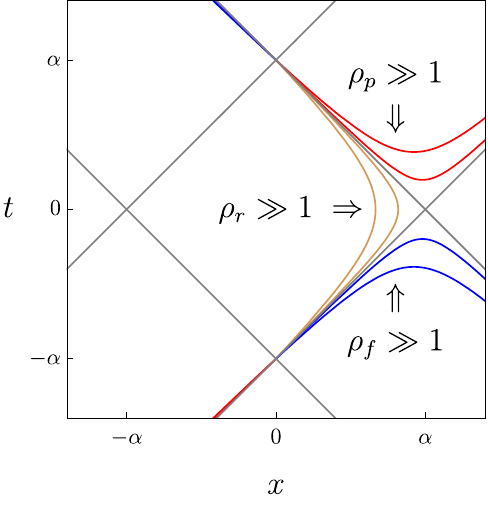}
     \caption{}\label{nhapfig2}
   \end{subfigure}
  \caption{Constant-$\rho$ trajectories for (a) $\rho_{r,p,f} \ll 1$, and (b) $\rho_{r,p,f} \gg 1$.}\label{nhapfig245}
\end{figure}

Based on the coordinate transformations between points in Minkowski spacetime restricted to the regions $D_R$, $D_P$, and $D_F$, and the causal diamond geometry associated to these regions, as given in Eqs.~\eqref{tdrxdretarho}, \eqref{tdrxdretarhodff1}–\eqref{tdrxdretarhodff}, we observe a characteristic behavior of constant-$\rho$ trajectories within these regions. Specifically, for small values of $\rho$, the trajectories asymptotically approach the left boundary of the causal diamond $D_R$, while for large values of $\rho$, they approach the right boundary. These features are illustrated in Fig.~\ref{nhapfig245}.

In order to characterize the boundary of the region $D_R$, let us begin by analyzing the expansion of $(t_{d_R}, x_{d_R})$ in the regime $\rho \ll 1$, with $\eta \in (-\infty, \infty)$, which is given by:
\begin{eqnarray}
t_{d_r} &=& 4 \rho \sinh\left(\frac{2 \eta}{\alpha}\right) - \frac{8 \rho^2 \sinh\left(\frac{4 \eta}{\alpha}\right)}{\alpha} + \frac{16 \rho^3 \sinh\left(\frac{6 \eta}{\alpha}\right)}{\alpha^2} + \mathcal{O}\qty[\qty(\rho/\alpha)^4],\label{r4rho1}
\end{eqnarray}
\begin{eqnarray}
x_{d_r} &=& -\alpha + 4 \rho \cosh\left(\frac{2 \eta}{\alpha}\right) - \frac{8 \rho^2 \cosh\left(\frac{4 \eta}{\alpha}\right)}{\alpha} + \frac{16 \rho^3 \cosh\left(\frac{6 \eta}{\alpha}\right)}{\alpha^2} + \mathcal{O}\qty[\qty(\rho/\alpha)^4].\label{r4rho2}
\end{eqnarray}

\begin{figure}[h]
\centering
\includegraphics[width=0.35\linewidth]{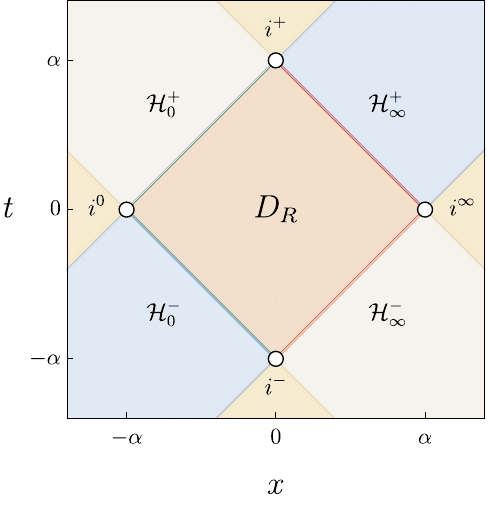}
\caption{Causal characterization of the boundary of the diamond $D_R$.}\label{figp4}
\end{figure}

In addition, since $\sinh(2\eta/\alpha) \approx \pm \cosh(2\eta/\alpha)$ for $\eta \gg 1$ and $\eta \ll -1$, respectively, it follows from Eqs.~\eqref{r4rho1} and \eqref{r4rho2} that, in the simultaneous limits $\rho \to 0$ and $\eta \to \pm\infty$, one obtains the following causal horizons:
\begin{eqnarray}
\mathcal{H}^+_0 &: \quad t_{d_r} = x_{d_r} + \alpha \,, \qquad \rho \to 0,\; \eta \to +\infty,
\end{eqnarray}
\begin{eqnarray}
\mathcal{H}^-_0 &: \quad t_{d_r} = -x_{d_r} - \alpha \,, \qquad \rho \to 0,\; \eta \to -\infty,
\end{eqnarray}
where $-\alpha < x_{d_r} < 0$. These horizons are depicted on the left boundary (blue) of the region $D_R$ in Fig.~\ref{figp4}.\\

On the other hand, we now turn to the expansion of $(t_{d_r}, x_{d_r})$ in the regime $\rho \gg 1$, with $\eta \in (-\infty, \infty)$, from which we obtain:
\begin{eqnarray}
t_{d_r} = \frac{\alpha^2 \sinh\left(\frac{2 \eta}{\alpha}\right)}{\rho} - \frac{\alpha^3 \sinh\left(\frac{4 \eta}{\alpha}\right)}{2 \rho^2} + \frac{\alpha^4 \sinh\left(\frac{6 \eta}{\alpha}\right)}{4 \rho^3} + \mathcal{O}\qty[\qty(\rho/\alpha)^{-4}],\label{rhoinf1}
\end{eqnarray}
\begin{eqnarray}
x_{d_r} = \alpha - \frac{\alpha^2 \cosh\left(\frac{2 \eta}{\alpha}\right)}{\rho} + \frac{\alpha^3 \cosh\left(\frac{4 \eta}{\alpha}\right)}{2 \rho^2} - \frac{\alpha^4 \cosh\left(\frac{6 \eta}{\alpha}\right)}{4 \rho^3} + \mathcal{O}\qty[\qty(\rho/\alpha)^{-4}].\label{rhoinf2}
\end{eqnarray}
Alternatively, the same expansion can be expressed in terms of the variable $\tilde{\rho}$, defined by $4\tilde{\rho} = \alpha^2 / \rho$, as
\begin{eqnarray}
t_{d_r} = 4\tilde{\rho}\sinh\left(\frac{2 \eta}{\alpha}\right) - \frac{8\tilde{\rho}^2 \sinh\left(\frac{4 \eta}{\alpha}\right)}{\alpha} + \frac{16\tilde{\rho}^3 \sinh\left(\frac{6 \eta}{\alpha}\right)}{\alpha^2} + \mathcal{O}\qty[\qty(\tilde{\rho}/\alpha)^{4}],\label{rhoinf1xx2}
\end{eqnarray}
\begin{eqnarray}
x_{d_r} = \alpha - 4\tilde{\rho}\cosh\left(\frac{2 \eta}{\alpha}\right) + \frac{8\tilde{\rho}^2 \cosh\left(\frac{4 \eta}{\alpha}\right)}{\alpha} - \frac{16\tilde{\rho}^2 \cosh\left(\frac{6 \eta}{\alpha}\right)}{\alpha^2} + \mathcal{O}\qty[\qty(\tilde{\rho}/\alpha)^{4}].\label{rhoinf2xx2}
\end{eqnarray}

As in the previous case, applying the limits $\rho \to \infty$ and $\eta \to \pm\infty$ to Eqs.~\eqref{rhoinf1} and \eqref{rhoinf2} yields the following causal horizons:
\begin{eqnarray}
\mathcal{H}^+_\infty &:& \quad t_{d_r} = -x_{d_r} + \alpha \quad,\quad \rho\to\infty\;,\;\eta \to \infty,
\end{eqnarray}
\begin{eqnarray}
\mathcal{H}^-_\infty &:& \quad t_{d_r} = x_{d_r} - \alpha \quad,\quad \rho\to\infty\;,\;\eta \to -\infty,
\end{eqnarray}
where $0 < x_{d_r} < \alpha$. These horizons are depicted on the right boundary (red) of the region $D_R$ in Fig.~\ref{figp4}.\\

Finally, the asymptotic regions of the causal diamond spacetime, corresponding to the four corners of $D_R$, can be identified by analyzing the coordinate transformations between Minkowski and causal diamond spacetimes in the appropriate limits:
\begin{eqnarray}
i^0 &:& \quad (t_{d_R},x_{d_R}) = (0,-\alpha) \quad,\quad \rho \to 0,\ \eta \in (-\infty,\infty),\\
i^\infty &:& \quad (t_{d_R},x_{d_R}) = (0,\alpha) \quad,\quad \rho \to \infty,\ \eta \in (-\infty,\infty),\\
i^- &:& \quad (t_{d_R},x_{d_R}) = (-\alpha,0) \quad,\quad \rho \in (0,\infty),\ \eta \to -\infty,\\
i^+ &:& \quad (t_{d_R},x_{d_R}) = (\alpha,0) \quad,\quad \rho \in (0,\infty),\ \eta \to \infty.
\end{eqnarray}

In Fig.~\ref{figp4}, we have included all the elements that characterize the causal structure of the boundary of the diamond $D_R$.

\section{Thermal behavior of causal horizons: Insights from causally disconnected observers}

\subsection{Method description}
The method implemented in the following subsections can be summarized in the application of the following steps:
\begin{itemize}
    \item[1.] The calculation of the Feynman propagator $G_F$ between events within causally disconnected regions of Minkowski spacetime, given by the right Rindler wedge $R$ and the causal diamond $D_R$, as well as between events whose spacetime interval crosses the past and future horizons of these regions.
    \item[2.] The application of the Fourier transform to the Feynman propagator in the interval $\Delta \eta$, which produces the frequency-space propagator $\tilde{G}_F$. This computation requires additional careful analysis for events whose spacetime interval crosses the causal horizons, due to the ill-defined notion of time associated with the $\eta$ coordinate in the regions $P$, $F$, $D_P$, and $D_F$.
    \item[3.] Implementation of the following setup: we consider two spacetime intervals in Minkowski, where $x^\mu$ is the common event between both intervals, $y^\mu$ lies in the past of $x^\mu$, while $\bar{y}^\mu$ lies in its future. Furthermore, $y^\mu$ and $\bar{y}^\mu$ are related by a reflection in the Minkowski time coordinate. This setup finally allows us to compute the ratio between the frequency-space propagators $\tilde{G}_F$ associated with the mentioned spacetime intervals.
\end{itemize}

\subsection{Thermality of the causal diamond horizons}
\begin{figure}[h]
\centering
\includegraphics[width=0.4\linewidth]{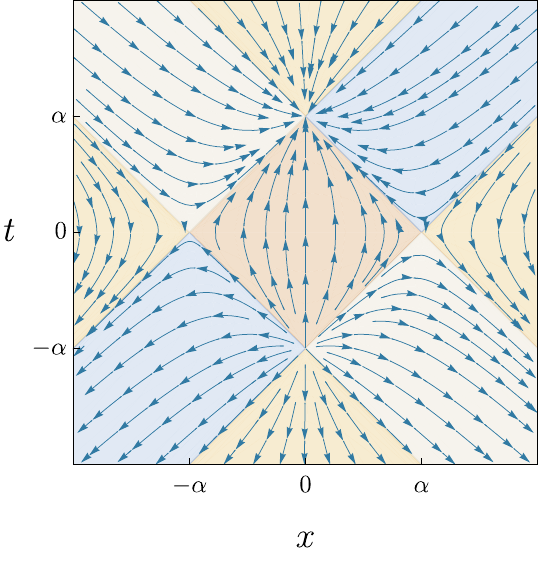}
\caption{Geometric flow associated with the Killing vector field $\partial_\eta$, which is timelike within the regions $D_R$ and $D_L$, and spacelike in the regions $D_F$ and $D_P$.}\label{kvcd}
\end{figure}

In this section, we analyze the propagation process from the perspective of an observer with finite lifetime $2\alpha$, confined to the region $D_R$, as illustrated in Fig.~\ref{cfcp}. The causal diamond spacetime metric, given by the line element in Eq.~\eqref{cdstm1}, explicitly depends on the temporal coordinate $\eta$, which precludes the existence of a Killing vector field associated with it. Nevertheless, the generator of temporal evolution, $\partial_\eta$, is a conformal Killing vector (the relevance of this vector field within the causal diamond geometry has been explored in Refs.~\cite{camblong2024conformal,chakraborty2024path,arzano2020conformal,camblong2023spectral}). The flow generated by $\partial_\eta$ in the extended Minkowski geometry, shown in Fig.~\ref{kvcd}, determines the coordinate transformations for the regions $D_L$, $D_P$, and $D_F$, as given in Eqs. \eqref{minktdl}–\eqref{tdrxdretarhodff}. Notably, $\partial_\eta$ is timelike in $D_R$ and $D_L$, with the temporal direction in $D_R$ aligned with Minkowski time, while it is reversed in $D_L$. This vector field becomes null on the causal horizons and is spacelike in the regions $D_P$ and $D_F$. Hence, the causal horizons precisely delimit the nature of $\partial_\eta$ and, consequently, the coordinate mapping in each region.\\

It is important to emphasize that, due to the explicit $\eta$-dependence of the causal diamond metric (this spacetime is not static), the full calculation of the frequency-space propagator $\tilde{G}_F$ for propagations within the region $D_R$, as well as for those crossing the past and future causal horizons, becomes significantly non-trivial. Nevertheless, applying a near-horizon analysis in the diamond region, we find that in the asymptotic regimes $\rho \ll 1$ and $\rho \gg 1$, the effective metric is static to first order in $\rho$ and $1/\rho$, respectively. Therefore, to simplify the analysis and properly define $\tilde{G}_F$, we restrict our attention to the regions near the horizons. This approach is natural, as the method does not impose any restriction on how close or far one must be from the horizon. In fact, for propagations that cross the causal horizons, what matters most is the crossing itself.

\subsubsection{Two propagations within the region $D_R$}

\begin{figure}[h]
 \centering
  \begin{subfigure}{0.3\textwidth}
    \includegraphics[width=\linewidth]{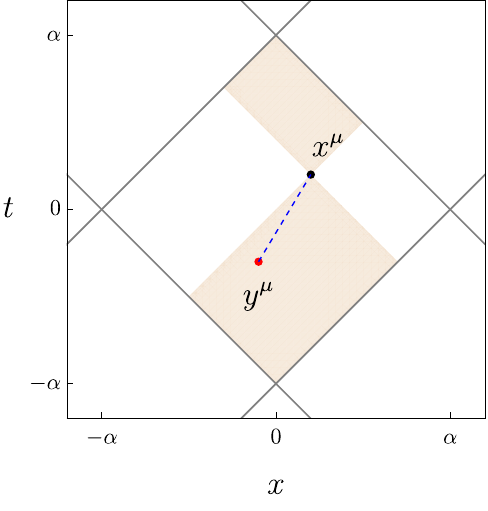}
    \caption{}\label{gpxy1d}
  \end{subfigure}
  \hspace{0.5cm}
  \begin{subfigure}{0.3\textwidth}
    \includegraphics[width=\linewidth]{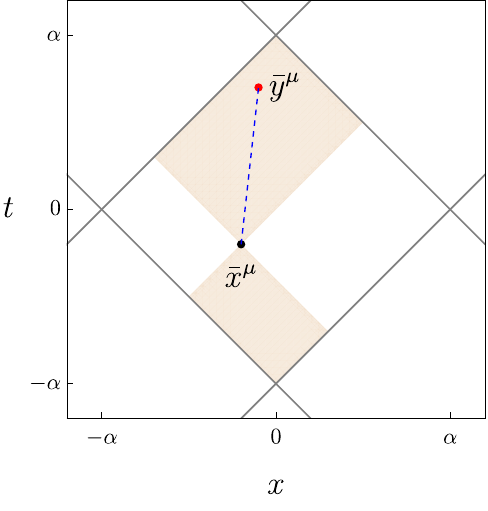}
    \caption{}\label{gpxy2d}
  \end{subfigure}
  \hspace{0.5cm}
  \begin{subfigure}{0.3\textwidth}
    \includegraphics[width=\linewidth]{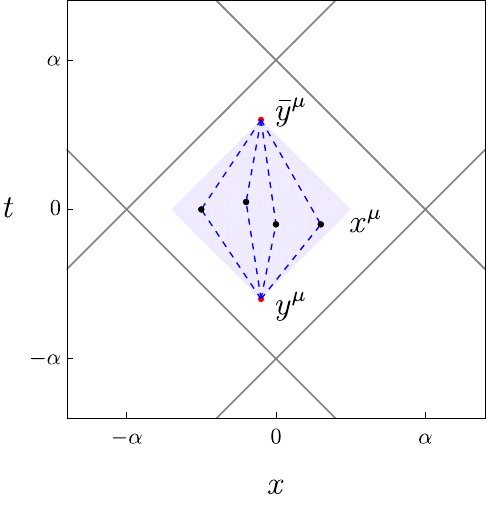}
    \caption{}\label{167}
  \end{subfigure}
  \caption{(a) Path~1: propagation from $y^\mu \in D_R$ to $x^\mu \in D_R$. (b) Path~2: propagation from $\bar{x}^\mu \in D_R$ to $\bar{y}^\mu \in D_R$. (c) Combined representation of Paths~1 and~2, with $x^\mu = \bar{x}^\mu$ and $y^\mu$ related to $\bar{y}^\mu$ by a time reflection.}
\end{figure}

We now consider the configurations depicted in Figs.~\ref{gpxy1d} and~\ref{gpxy2d}. Path~1 corresponds to the propagation from the event $y^\mu \equiv (y^0, y^1)$ to $x^\mu \equiv (x^0, x^1)$, while Path~2 corresponds to the propagation from $\bar{x}^\mu \equiv (\bar{x}^0, \bar{x}^1)$ to $\bar{y}^\mu \equiv (\bar{y}^0, \bar{y}^1)$, both confined within the region $D_R$. To analyze these processes from the viewpoint of the causal–diamond observer, we employ the coordinate transformation given in Eq.~\eqref{tdrxdretarho}:
\begin{eqnarray}
x^0 &=& \frac{4\alpha^2\rho_x\sinh(\tfrac{2\eta_x}{\alpha})}{\alpha^2 + 4\alpha\rho_x\cosh(\tfrac{2\eta_x}{\alpha}) + 4\rho_x^2}, \qquad
x^1 = \frac{4\alpha\rho_x^2 - \alpha^3}{\alpha^2 + 4\alpha\rho_x\cosh(\tfrac{2\eta_x}{\alpha}) + 4\rho_x^2},
\end{eqnarray}
\begin{eqnarray}
y^0 &=& \frac{4\alpha^2\rho_y\sinh(\tfrac{2\eta_y}{\alpha})}{\alpha^2 + 4\alpha\rho_y\cosh(\tfrac{2\eta_y}{\alpha}) + 4\rho_y^2}, \qquad
y^1 = \frac{4\alpha\rho_y^2 - \alpha^3}{\alpha^2 + 4\alpha\rho_y\cosh(\tfrac{2\eta_y}{\alpha}) + 4\rho_y^2},
\end{eqnarray}

\begin{eqnarray}
\bar{x}^0 &=& \frac{4\alpha^2\bar{\rho}_x\sinh(\tfrac{2\bar{\eta}_x}{\alpha})}{\alpha^2 + 4\alpha\bar{\rho}_x\cosh(\tfrac{2\bar{\eta}_x}{\alpha}) + 4\bar{\rho}_x^2}, \qquad
\bar{x}^1 = \frac{4\alpha\bar{\rho}_x^2 - \alpha^3}{\alpha^2 + 4\alpha\bar{\rho}_x\cosh(\tfrac{2\bar{\eta}_x}{\alpha}) + 4\bar{\rho}_x^2},
\end{eqnarray}
\begin{eqnarray}
\bar{y}^0 &=& \frac{4\alpha^2\bar{\rho}_y\sinh(\tfrac{2\bar{\eta}_y}{\alpha})}{\alpha^2 + 4\alpha\bar{\rho}_y\cosh(\tfrac{2\bar{\eta}_y}{\alpha}) + 4\bar{\rho}_y^2}, \qquad
\bar{y}^1 = \frac{4\alpha\bar{\rho}_y^2 - \alpha^3}{\alpha^2 + 4\alpha\bar{\rho}_y\cosh(\tfrac{2\bar{\eta}_y}{\alpha}) + 4\bar{\rho}_y^2}.
\end{eqnarray}

From these expressions we obtain the squared spacetime intervals:
\begin{eqnarray}
\text{Path~1:}\quad\sigma^2(x,y) &=& -(x^0-y^0)^2+(x^1-y^1)^2\nonumber\\
&&\hspace{-1.5cm}= \frac{16 \alpha^4 \left[\rho_{x}^2 + \rho_{y}^2 - 2 \rho_{x} \rho_{y} \cosh\!\left( \tfrac{2 (\eta_{x} - \eta_{y})}{\alpha} \right) \right]}{\left[ \alpha^2 + 4 \alpha \rho_{x} \cosh\!\left( \tfrac{2 \eta_{x} }{\alpha} \right) + 4 \rho_{x}^2 \right] \left[ \alpha^2 + 4 \alpha \rho_{y} \cosh\!\left( \tfrac{2 \eta_{y} }{\alpha} \right) + 4 \rho_{y}^2 \right]},\label{itdrdr122}
\end{eqnarray}
\begin{eqnarray}
\text{Path~2:}\quad\sigma^2(\bar{x},\bar{y}) &=& -(\bar{y}^0-\bar{x}^0)^2+(\bar{y}^1-\bar{x}^1)^2\nonumber\\
&&\hspace{-1.5cm}= \frac{16 \alpha^4 \left[\bar{\rho}_{x}^2 + \bar{\rho}_{y}^2 - 2 \bar{\rho}_{x} \bar{\rho}_{y} \cosh\!\left( \tfrac{2 (\bar{\eta}_{x} - \bar{\eta}_{y})}{\alpha} \right) \right]}{\left[ \alpha^2 + 4 \alpha \bar{\rho}_{x} \cosh\!\left( \tfrac{2 \bar{\eta}_{x} }{\alpha} \right) + 4 \bar{\rho}_{x}^2 \right] \left[ \alpha^2 + 4 \alpha \bar{\rho}_{y} \cosh\!\left( \tfrac{2 \bar{\eta}_{y} }{\alpha} \right) + 4 \bar{\rho}_{y}^2 \right]}.\label{itdrdr12}
\end{eqnarray}

As expected, due to the explicit $\eta$–dependence of the causal–diamond metric, the temporal dependence in Eqs.~\eqref{itdrdr122} and~\eqref{itdrdr12} arises not only through the differences $\Delta\eta = \eta_x - \eta_y$ and $\Delta\bar{\eta} = \bar{\eta}_x - \bar{\eta}_y$, but also through the individual coordinates $\eta_{x,y}$ and $\bar{\eta}_{x,y}$ appearing in the denominators. To simplify the analysis, we therefore work in the near–horizon approximation, where to leading order in the expansions given by Eqs.~\eqref{r4rho1} and~\eqref{r4rho2} for $\rho \ll 1$, and by Eqs.~\eqref{rhoinf1} and~\eqref{rhoinf2} for $\rho \gg 1$, the spacetime geometry becomes effectively static:

\begin{itemize}
\item[(1)] For $\rho \ll 1$:
\begin{eqnarray}
\text{Path~1:}\quad\sigma^2(x,y) &=& -(x^0 - y^0)^2 + (x^1 - y^1)^2 \nonumber\\
&&\hspace{-2cm}= 16\rho_{x}^2 + 16\rho_{y}^2 - 32\rho_{x}\rho_{y}\cosh\!\qty[\frac{2(\eta_{x} - \eta_{y})}{\alpha}] + \mathcal{O}(\rho^3_{x}, \rho^3_{y}),\label{eqcdsspt1}
\end{eqnarray}
\begin{eqnarray}
\text{Path~2:}\quad\sigma^2(\bar{x}, \bar{y}) &=& -(\bar{y}^0 - \bar{x}^0)^2 + (\bar{y}^1 - \bar{x}^1)^2 \nonumber\\
&&\hspace{-2cm}= 16\bar{\rho}^2_{x} + 16\bar{\rho}^2_{y} - 32\bar{\rho}_{x}\bar{\rho}_{y}\cosh\!\qty[\frac{2(\bar{\eta}_{x} - \bar{\eta}_{y})}{\alpha}] + \mathcal{O}(\bar{\rho}^3_{x}, \bar{\rho}^3_{y}).\label{eqcdsspt2}
\end{eqnarray}

\item[(2)] For $\rho \gg 1$, or equivalently $\tilde{\rho} \ll 1$, by means of
$4\tilde{\rho} = \alpha^2 / \rho$:
\begin{eqnarray}
\text{Path~1:}\quad\sigma^2(x,y) &=& -(x^0 - y^0)^2 + (x^1 - y^1)^2 \nonumber\\
&&\hspace{-2cm}= \frac{\alpha^4}{\rho^2_{x}} + \frac{\alpha^4}{\rho^2_{y}} - \frac{2\alpha^4}{\rho_{x}\rho_{y}}\cosh\!\qty[\frac{2(\eta_{x} - \eta_{y})}{\alpha}] + \mathcal{O}(\rho^{-3}_{x}, \rho^{-3}_{y}),\label{bash12}\\
&&\hspace{-2cm}= 16\tilde{\rho}^2_{x} + 16\tilde{\rho}^2_{y} - 32\tilde{\rho}_{x}\tilde{\rho}_{y}\cosh\!\qty[\frac{2(\eta_{x} - \eta_{y})}{\alpha}] + \mathcal{O}(\tilde{\rho}^3_{x}, \tilde{\rho}^3_{y}),\label{eqcdsspt3}
\end{eqnarray}
\begin{eqnarray}
\text{Path~2:}\quad\sigma^2(\bar{x}, \bar{y}) &=& -(\bar{y}^0 - \bar{x}^0)^2 + (\bar{y}^1 - \bar{x}^1)^2 \nonumber\\
&&\hspace{-2cm}= \frac{\alpha^4}{\bar{\rho}^2_{x}} + \frac{\alpha^4}{\bar{\rho}^2_{y}} - \frac{2\alpha^4}{\bar{\rho}_{x}\bar{\rho}_{y}}\cosh\!\qty[\frac{2(\bar{\eta}_{x} - \bar{\eta}_{y})}{\alpha}] + \mathcal{O}(\bar{\rho}^{-3}_{x}, \bar{\rho}^{-3}_{y}),\\
&&\hspace{-2cm}= 16\tilde{\bar{\rho}}^2_{x} + 16\tilde{\bar{\rho}}^2_{y} - 32\tilde{\bar{\rho}}_{x}\tilde{\bar{\rho}}_{y}\cosh\!\qty[\frac{2(\bar{\eta}_{x} - \bar{\eta}_{y})}{\alpha}] + \mathcal{O}(\tilde{\bar{\rho}}^3_{x}, \tilde{\bar{\rho}}^3_{y}).\label{bash122}
\end{eqnarray}
\end{itemize}

We emphasize the following observations valid in both regimes, $\rho \ll 1$ and $\rho \gg 1$ (expressed in terms of $\tilde{\rho} \ll 1$): (1) The functional forms of the squared spacetime intervals corresponding to Paths~1 and~2 are manifestly identical, up to a simple relabeling of variables. This equivalence arises because both propagations take place within the same region, namely $D_R$, and therefore share the same mapping between causal–diamond and Minkowski coordinates. (2) The time dependence appears solely through the intervals $\Delta\eta = \eta_{x} - \eta_{y}$ and $\Delta\bar{\eta} = \bar{\eta}_{x} - \bar{\eta}_{y}$, as a direct consequence of the static character of the effective spacetime geometry. This property allows one to define the corresponding frequency–space propagators via Fourier transforms with respect to $\Delta\eta$ and $\Delta\bar{\eta}$, respectively, as introduced in Eq.~\eqref{foutregf2}. (3) The squared spacetime intervals for both Paths~1 and~2 are invariant under the transformations $\Delta\eta \to -\Delta\eta$ and $\Delta\bar{\eta} \to -\Delta\bar{\eta}$, respectively. This invariance is inherited from the inertial propagator and ultimately reflects the fact that the coordinate transformation relates two static geometries—Minkowski space and the near-horizon approximation of the causal-diamond geometry—which, to leading order, reduces to the Rindler form, as seen from Eqs.~\eqref{r4rho1}–\eqref{r4rho2} and \eqref{rhoinf1xx2}–\eqref{rhoinf2xx2}.\\

To compare Paths~1 and~2, we adopt the setup $x^\mu = \bar{x}^\mu$ and $\bar{y}^\mu = (-y^0, y^1)$, as shown in Fig.~\ref{167}. Under this configuration, for $\rho \ll 1$ and neglecting higher–order terms in~$\rho$, we obtain
\begin{eqnarray}
\text{Path~1:}\quad\sigma^2(x,y) &\approx& 16\rho_{x}^2 + 16\rho_{y}^2 - 32\rho_{x} \rho_{y} \cosh\!\qty[\tfrac{2(\eta_{x} - \eta_{y})}{\alpha}],\label{eqcdsspt5}
\end{eqnarray}
\begin{eqnarray}
\text{Path~2:}\quad\sigma^2(x,\bar{y}) &\approx& 16\rho_{x}^2 + 16\rho_{y}^2 - 32\rho_{x} \rho_{y} \cosh\!\qty[\tfrac{2(\eta_{x} + \eta_{y})}{\alpha}],\label{eqcdsspt6}
\end{eqnarray}
where the setup implies $\bar{\eta}_x=\eta_x$, $\bar{\rho}_x=\rho_x$, $\bar{\eta}_y=-\eta_y$, and $\bar{\rho}_y=\rho_y$. Consequently, $\Delta\eta=\eta_x-\eta_y$ and $\Delta\bar{\eta}=\bar{\eta}_x-\bar{\eta}_y=\eta_x+\eta_y$. For $\rho\gg1$ (expressed through $\tilde{\rho}$), the structure of Eqs.~\eqref{eqcdsspt5}–\eqref{eqcdsspt6} remains identical, with the replacement $\rho \to \tilde{\rho}$. Hence, it suffices to restrict our analysis to the near–horizon regime $\rho \ll 1$, since the other asymptotic case is completely analogous.\\

The frequency–space propagator associated with Path~1 is then
\begin{eqnarray}
\tilde{G}_F^{\text{Path~1}}(\Omega, \rho_{x}, \rho_{y}) 
&=& \int_{-\infty}^{\infty} d(\Delta \eta)\, G_F^{\text{Path~1}}(\Delta\eta, \rho_{x}, \rho_{y})\, e^{-i\Omega \Delta\eta} \nonumber\\
&=& \int_0^\infty \frac{ds}{4\pi s} \exp\!\qty[\frac{4i}{s}(\rho_{x}^2 + \rho_{y}^2) - i s m^2 - s \epsilon]\times\nonumber\\
&&\times\int_{-\infty}^\infty d(\Delta \eta)\, \exp\!\qty[-\frac{8i}{s} \rho_{x} \rho_{y} \cosh(\tfrac{2\Delta \eta}{\alpha})] e^{-i\Omega \Delta \eta} \nonumber\\
&=& \int_0^\infty \frac{ds}{4\pi s} \exp\!\qty[\frac{4i}{s}(\rho_{x}^2 + \rho_{y}^2) - i s m^2 - s \epsilon] 
\qty[\alpha K_{i\alpha\Omega/2}\!\qty(\tfrac{8i \rho_{x} \rho_{y}}{s})]. \label{ataumnor}
\end{eqnarray}

Here we have employed Eq.~$7^7$ of Sec.~8.432 in Ref.~\cite{gradshteyn2014table},
\begin{equation}
    K_\nu(\zeta z) = \frac{z^\nu}{2} \int_0^\infty \exp\!\qty[-\zeta \tfrac{\gamma + z^2 \gamma^{-1}}{2}]\, \gamma^{-\nu - 1} d\gamma, 
    \qquad |\arg(z)| \leq \frac{\pi}{4}, \qquad \Re(\nu) < 1,\label{gradeqeq}
\end{equation}
with the identifications $\zeta = 8i \rho_{x} \rho_{y}/s$, $\gamma = e^{2\Delta\eta/\alpha}$, $z = 1$, and $\nu = i\alpha\Omega / 2$.

Similarly, for Path~2 we have
\begin{equation}
    \tilde{G}_F^{\text{Path~2}}(\bar{\Omega}, \rho_{x}, \rho_{y})
    = \int_{-\infty}^{\infty} d(\Delta \bar{\eta})\, G_F^{\text{Path~2}}(\Delta\bar{\eta}, \rho_{x}, \rho_{y})\, e^{-i\bar{\Omega} \Delta\bar{\eta}}.
\end{equation}
Since $\Delta\eta$ and $\Delta\bar{\eta}$ are dummy integration variables, it is clear that by choosing $\bar{\Omega} = \Omega$ the two expressions become identical. Thus we obtain
\begin{equation}
 \tilde{G}_F^{\text{Path~1}}(\Omega, \rho_{x}, \rho_{y}) 
 = \tilde{G}_F^{\text{Path~2}}(\Omega, \rho_{x}, \rho_{y}).\label{case1path12R}
\end{equation}

This result shows that the propagations along Paths~1 and~2, corresponding to event pairs within the region $D_R$, where the time evolution generated by $\partial_\eta$ is well defined and hence so is the interval $\Delta\eta$, are physically equivalent. This is fully consistent with the expectation from the viewpoint of an inertial observer in Minkowski spacetime.

\subsubsection{Propagation from the region $D_P$ to $D_R$, and from $D_R$ to $D_F$}
\begin{figure}[h]
 \centering
   \begin{subfigure}{0.4\textwidth}
    \includegraphics[width=\linewidth]{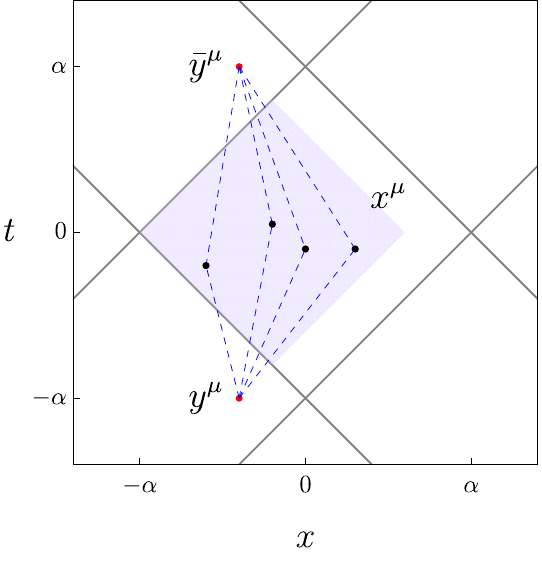}
     \caption{}\label{gpxy11d}
   \end{subfigure}
   \hspace{0.5cm}
   \begin{subfigure}{0.4\textwidth}
     \centering
     \includegraphics[width=\linewidth]{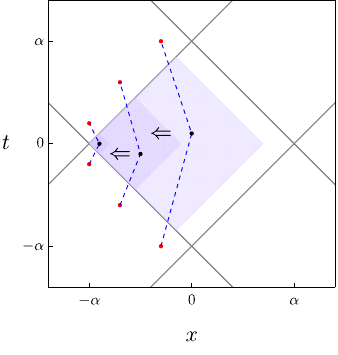}
     \caption{}\label{nhcd12}
   \end{subfigure}
  \caption{(a) Paths 1 and 2, where $x^\mu = \bar{x}^\mu$ and $y^\mu$ is related to $\bar{y}^\mu$ by a time reflection.\hspace{1cm}(b) Near-horizon approximation.}
\end{figure}

We now proceed with the configuration shown in Fig.~\ref{gpxy11d}, where Path~1 corresponds to propagation from the event $y^\mu \equiv (y^0, y^1) \in D_P$ to $x^\mu \equiv (x^0, x^1) \in D_R$, and Path~2 corresponds to propagation from $x^\mu \equiv (x^0, x^1) \in D_R$ to $\bar{y}^\mu \equiv (\bar{y}^0, \bar{y}^1) \in D_F$. We impose the identifications $\bar{y}^0 = -y^0$ and $\bar{y}^1 = y^1$, from which it follows that $\bar{\rho}_f = \rho_p$ and $\bar{\eta}_f = -\eta_p$. Accordingly, the computation of the squared spacetime interval requires only the coordinate transformations given by Eqs.~\eqref{tdrxdretarho} and~\eqref{tdrxdretarhodff}:
\begin{eqnarray}
x^0 = \frac{4\alpha^2\rho_r\sinh(\frac{2\eta_r}{\alpha})}{\alpha^2 + 4\alpha\rho_r\cosh(\frac{2\eta_r}{\alpha}) + 4\rho^2_r}\quad&, &\quad
x^1 = \frac{-4\alpha\rho^2_f - \alpha^3}{\alpha^2 + 4\alpha\rho_f\sinh(\frac{2\eta_f}{\alpha}) - 4\rho^2_f},
\end{eqnarray}
\begin{eqnarray}
y^0 = \frac{-4\alpha^2\rho_p\cosh(\frac{2\eta_p}{\alpha})}{\alpha^2 - 4\alpha\rho_p\sinh(\frac{2\eta_p}{\alpha}) - 4\rho^2_p}\quad&, &\quad
y^1 = \frac{-4\alpha\rho^2_p - \alpha^3}{\alpha^2 - 4\alpha\rho_p\sinh(\frac{2\eta_p}{\alpha}) - 4\rho^2_p}.
\end{eqnarray}

Thus, we obtain the following squared spacetime intervals:
\begin{eqnarray}
\text{{Path 1:}}\quad\sigma^2(x,y)&=& -(x^0-y^0)^2+(x^1-y^1)^2\nonumber\\
&&\hspace{-2cm}=\frac{16 \alpha^4 \left(-2 \rho_{p} \rho_{r} \sinh\left(\frac{2(\eta_{r} - \eta_{p})}{\alpha}\right) - \rho_{p}^2 + \rho_{r}^2\right)}{\left(\alpha^2 - 4 \alpha \rho_{p} \sinh\left(\frac{2 \eta_{p}}{\alpha}\right) - 4 \rho_{p}^2\right)\left(\alpha^2 + 4 \alpha \rho_{r} \cosh\left(\frac{2 \eta_{r}}{\alpha}\right) + 4 \rho_{r}^2\right)},\label{screcd12}
\end{eqnarray}
\begin{eqnarray}
\text{{Path 2:}}\quad\sigma^2(x,\bar{y})&=& -(\bar{y}^0-x^0)^2+(\bar{y}^1-x^1)^2\nonumber\\
&&\hspace{-2cm}=\frac{16 \alpha^4 \left(2 \rho_{p} \rho_{r} \sinh\left(\frac{2(\eta_{r} + \eta_{p})}{\alpha}\right) - \rho_{p}^2 + \rho_{r}^2\right)}{\left(\alpha^2 - 4 \alpha \rho_{p} \sinh\left(\frac{2 \eta_{p}}{\alpha}\right) - 4 \rho_{p}^2\right)\left(\alpha^2 + 4 \alpha \rho_{r} \cosh\left(\frac{2 \eta_{r}}{\alpha}\right) + 4 \rho_{r}^2\right)}.\label{screcd22}
\end{eqnarray}

As seen in Eqs.~\eqref{screcd12} and \eqref{screcd22}, the dependence on $\eta_{r,p}$ is intricate. However, the analysis can be simplified by considering the near-horizon approximation, where the spacetime geometry becomes effectively static. In this configuration (see Fig.~\ref{nhcd12}), the boundaries between $D_P$ and $D_R$, and between $D_R$ and $D_F$, correspond to the left past horizon $\mathcal{H}^-_0$ and the left future horizon $\mathcal{H}^+_0$, respectively. By approaching these boundaries via the approximations $\rho_{r,p,f} \ll 1$, as illustrated in Fig.~\ref{nhapfig}, we obtain:

\begin{eqnarray}
\text{{Path 1:}}\quad\sigma^2(x,y)&=&  -(x^0-y^0)^2+(x^1-y^1)^2\nonumber\\
&=& 16\rho_r^2 - 16\rho_p^2 - 32\rho_p\rho_r \sinh\qty[\frac{2(\eta_r - \eta_p)}{\alpha}] + \mathcal{O}(\rho^3)\nonumber\\
&\approx& -32\rho_p\rho_r \sinh\qty[\frac{2(\eta_r - \eta_p)}{\alpha}],\\
\text{{Path 2:}}\quad\sigma^2(x,\bar{y})&=& -(\bar{y}^0-x^0)^2+(\bar{y}^1-x^1)^2\nonumber\\
&=& 16\rho_r^2 - 16\rho_p^2 + 32\rho_p\rho_r \sinh\qty[\frac{2(\eta_r + \eta_p)}{\alpha}] + \mathcal{O}(\rho^3)\nonumber\\
&\approx& 32\rho_p\rho_r \sinh\qty[\frac{2(\eta_r + \eta_p)}{\alpha}],
\end{eqnarray}
where we have assumed, without loss of generality within the present approximation, that $16\rho_r^2 - 16\rho_p^2 \approx 0$.\\

As we can see, the propagations for Paths~1 and~2 occur between regions with different coordinate charts, where the notions of time and space are effectively interchanged. As a result, the squared spacetime intervals for these paths are manifestly different, and they also differ from the case in which both Path~1 and Path~2 lie within the same region $D_R$, as given in Eqs.~\eqref{eqcdsspt5} and \eqref{eqcdsspt6}, where the temporal-interval-dependent function was expressed in terms of the hyperbolic cosine. Nevertheless, the time dependence continues to appear through the intervals $\Delta\eta = \eta_r - \eta_p$ and $\Delta\bar{\eta} = \eta_r - \bar{\eta}_f=\eta_r+\eta_p$, since in the regions relevant to this analysis the effective geometry remains independent of $\eta$. Moreover, in the present case, the squared spacetime intervals for Paths~1 and~2 are no longer invariant under the transformations $\Delta\eta \to -\Delta\eta$ and $\Delta\bar{\eta} \to - \Delta\bar{\eta}$, respectively. This clearly indicates not only that the two propagations are physically distinct, but also reflects the interchange of time and space in the coordinate maps employed. In addition, because the notion of $\eta$ as a time coordinate is ill-defined in the $D_P$ and $D_F$ regions, these intervals are to be understood in such a way that $\eta_p$ and $\bar{\eta}_f$ — not being physically meaningful time coordinates — merely correspond to shifts of the Rindler time coordinate in region $D_R$. Consequently, the Fourier transforms of $G_F$ with respect to $\Delta\eta$ and $\Delta\bar{\eta}$ are effectively carried out in terms of $\eta_r$ and $\bar{\eta}_r$, respectively.\\

The frequency-space propagator in the near-horizon approximation for Path~1 is given by
\begin{eqnarray}
    \tilde{G}_F^{\text{Path 1}}(\Omega, \rho_p, \rho_r) &&= \int_{-\infty}^{\infty} d\eta_r\, G_F^{\text{Path 1}}(\eta_r-\eta_p,\rho_p, \rho_r)e^{-i\Omega\eta_r}\nonumber\\
&&= \int_{-\infty}^{\infty} d(\Delta \eta)\, G_F^{\text{Path 1}}(\Delta\eta,\rho_p, \rho_r)e^{-i\Omega \Delta\eta}e^{-i\Omega\eta_p}\nonumber\\
&=& \int_{-\infty}^{\infty} d(\Delta \eta) \qty(\frac{1}{4\pi}\int_0^\infty \frac{ds}{s} e^{\left[\frac{i}{4s}\qty(-32\rho_p\rho_r \sinh\qty[\frac{2\Delta \eta}{\alpha}]) - i s m^2 - s \epsilon\right]})e^{-i\Omega \Delta\eta}e^{-i\Omega\eta_p}\nonumber\\
&=& \int_0^\infty \frac{ds}{4\pi s} e^{- i s m^2 - s \epsilon}\int_{-\infty}^{\infty} d(\Delta \eta)\, e^{-\frac{8i}{s}\rho_p\rho_r \sinh\qty[\frac{2\Delta \eta}{\alpha}]}e^{-i\Omega \Delta\eta}e^{-i\Omega\eta_p}\nonumber\\
&=& \int_0^\infty \frac{ds}{4\pi s} e^{- i s m^2 - s \epsilon}\qty[\alpha e^{-i\Omega\eta_p}e^{-\frac{\pi\Omega\alpha}{4}}K_{-i\Omega\alpha/2}\qty(-\frac{8\rho_p\rho_r}{s})],\label{nhaaplus}
\end{eqnarray}
where we have used the identity
\begin{equation}
    2e^{-\operatorname{sign}(\zeta)\pi\nu/2}K_{i\nu}(2\zeta) = \int_0^\infty \exp\qty[i\zeta (\gamma - \gamma^{-1})] \gamma^{i\nu - 1} d\gamma, \quad \zeta\in\mathbb{R},\;-1<\Im(\nu)<1,\label{formulmaster}
\end{equation}
with the substitutions $\zeta = -4\rho_p \rho_r / s$, $\gamma = e^{2\Delta\eta/\alpha}$, and $\nu = -\Omega\alpha/2$. Thus, we have:
\begin{equation}
    \tilde{G}_F^{\text{Path 1}}(-\Omega, \rho_p, \rho_r) = e^{\pi\Omega\alpha/2}\, \tilde{G}_F^{\text{Path 1}}(\Omega, \rho_p, \rho_r).\label{bolztfa22}
\end{equation}

Next, for Path 2, we obtain:
\begin{eqnarray}
\tilde{G}_F^{\text{Path 2}}(\bar{\Omega}, \rho_p, \rho_r) &=& \int_{-\infty}^{\infty} d\eta_r\, G_F^{\text{Path 2}}(\eta_r+\eta_p,\rho_p, \rho_r)\, e^{-i\bar{\Omega} \eta_r}\nonumber\\
&=& \int_{-\infty}^{\infty} d(\Delta \bar{\eta})\, G_F^{\text{Path 2}}(\Delta\bar{\eta},\rho_p, \rho_r)\, e^{-i\bar{\Omega} \Delta\bar{\eta}}e^{i\bar{\Omega}\eta_p}\nonumber\\
&=& \int_{-\infty}^{\infty} d(\Delta \bar{\eta}) \qty(\frac{1}{4\pi}\int_0^\infty \frac{ds}{s} e^{\left[\frac{i}{4s}\qty(32\rho_p\rho_r \sinh\qty[\frac{2\Delta\bar{\eta}}{\alpha}]) - i s m^2 - s \epsilon\right]})e^{-i\bar{\Omega} \Delta\bar{\eta}}e^{i\bar{\Omega}\eta_p}\nonumber\\
&=& \int_0^\infty \frac{ds}{4\pi s} e^{- i s m^2 - s \epsilon} \int_{-\infty}^{\infty} d(\Delta \bar{\eta})\, e^{\frac{8i}{s}\rho_p\rho_r \sinh\qty[\frac{2\Delta\bar{\eta}}{\alpha}]} e^{-i\bar{\Omega} \Delta\bar{\eta}}e^{i\bar{\Omega}\eta_p}\nonumber\\
&=&\int_0^\infty \frac{ds}{4\pi s} e^{- i s m^2 - s \epsilon} \qty[\alpha e^{i\bar{\Omega}\eta_p}e^{\frac{\pi\bar{\Omega}\alpha}{4}}K_{i\bar{\Omega}\alpha/2}\qty(-\frac{8\rho_p\rho_r}{s})],
\end{eqnarray}
where, once again, we have used the change $\Delta\bar{\eta}\to - \Delta\bar{\eta}$, and the identity given in Eq.~\eqref{formulmaster}, with the substitutions $\zeta = -4\rho_p \rho_r / s$, $\gamma = e^{2\Delta\bar{\eta}/\alpha}$, and $\nu = \bar{\Omega}\alpha/2$.\\

Furthermore, from Eq.~\eqref{nhaaplus}, identifying $\bar{\Omega} = \Omega$, we obtain the following relation:
\begin{equation}
\tilde{G}_F^{\text{Path 2}}(\Omega, \rho_p, \rho_r) = \tilde{G}_F^{\text{Path 1}}(-\Omega, \rho_p, \rho_r).
\end{equation}

Hence, from Eq.~\eqref{bolztfa22} we obtain:
\begin{equation}
\abs{\frac{\tilde{G}_F^{\text{Path 1}}(\Omega, \rho_p, \rho_r)}{\tilde{G}_F^{\text{Path 2}}(\Omega, \rho_p, \rho_r)} }^2 = e^{-\beta\Omega} \quad,\quad \beta = \pi\alpha. \label{emiproscas4}
\end{equation}

Remarkably, this ratio takes the form of a thermal state, characterized by a Boltzmann factor, from which the effective temperature is given by 
\begin{equation}
T = \frac{1}{\pi \alpha}.
\end{equation}
The temperature obtained through this method is consistent with previous findings based on the thermal time hypothesis, the open quantum system approach, the thermofield-double construction, and related frameworks~\cite{martinetti2003diamond,chakraborty2022thermal,chakraborty2024path,camblong2024conformal,arzano2020conformal,su2016spacetime,foo2020generating,foo2025superpositions}.\\

Therefore, the propagation associated with Path~1 can be interpreted as the emission of scalar quanta from the past horizon $\mathcal{H}^-_0$, as seen from the perspective of the observer in $D_R$, while the propagation associated with Path~2 represents the absorption of these scalar quanta at the future horizon $\mathcal{H}^+_0$. Then, although propagation across causal horizons appears physically equivalent from the perspective of an inertial observer, restricting the observation to an observer confined to the causal diamond $D_R$ reveals the thermal nature of these horizons, a feature otherwise hidden in the inertial propagator.\\

It is also worth mentioning that an entirely analogous analysis can be performed near the right boundary of $D_R$, by noting that the regime $\rho \gg 1$ can be mapped to $\tilde{\rho} \ll 1$, rendering the whole treatment formally equivalent. In addition, for completeness, Sec.~\ref{apnaa1} presents the application of this method to the Rindler–wedge framework, which provides a causally consistent extension and broadens the physical interpretation of the results obtained in Ref.~\cite{padmanabhan2019thermality}.

\subsection{Thermality of the Rindler horizons}\label{apnaa1}
\begin{figure}[h]
\centering
\includegraphics[width=0.35\linewidth]{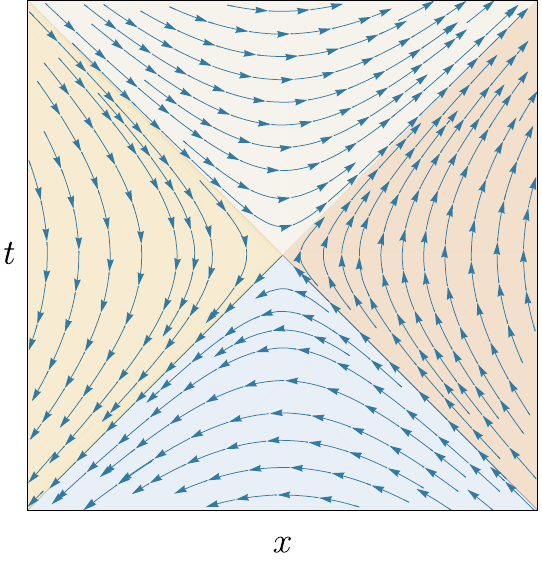}
\caption{Geometric flow associated with the Killing vector field $\partial_\eta$: timelike in wedges $R$ and $L$, spacelike in $P$ and $F$, and null on the horizons $t=\pm x$.}\label{kvr}
\end{figure}

As discussed in the case of the causal diamond, the Killing vector field $\partial_\eta$ plays a central role in determining the causal properties of the corresponding spacetime region. In the present section, we extend this analysis to the Rindler geometry, examining field propagation from the viewpoint of an observer undergoing uniform acceleration in Minkowski spacetime, whose trajectory is confined to the right Rindler wedge $R$ (see Fig.~\ref{rrw}). The observer’s temporal evolution is governed by the vector field $\partial_\eta$, whose geometric flow—depicted in Fig.~\ref{kvr} and defined by the coordinate map in Eq.~\eqref{trrhoeta1}—corresponds, up to a constant factor, to the Lorentz-boost generator in Minkowski space. The same flow extends naturally to the remaining wedges $L$, $F$, and $P$, with coordinate relations given in Eqs.~\eqref{trrhoeta2}–\eqref{trrhoeta4}.

As in the causal–diamond case, the character of $\partial_\eta$ changes upon crossing the causal horizons $t=\pm x$, thereby altering the interpretation of $\eta$ as a time coordinate. In the wedge $R$, $\partial_\eta$ is timelike and future-directed; on the horizons it becomes null; in $L$ it is timelike but past-directed; and in $P$ and $F$ it is spacelike, meaning that $\eta$ no longer represents a physical time parameter.\\

With these properties in mind, we now analyze the Feynman propagator $G_F$ for events whose intervals lie entirely within $R$, as well as those that cross the past and future Rindler horizons—specifically, propagation from $P$ into $R$ and from $R$ into $F$. The corresponding frequency–space propagator $\tilde{G}_F$ will then clarify how the causal horizon—marking the transition in the character of $\partial_\eta$ and, consequently, in the coordinate mapping—affects the ratio between propagations that are physically equivalent from the standpoint of an inertial observer.

\subsubsection{Two propagations in the $R$ wedge}
\begin{figure}[h]
 \centering
  \begin{subfigure}{0.3\textwidth}
    \includegraphics[width=\linewidth]{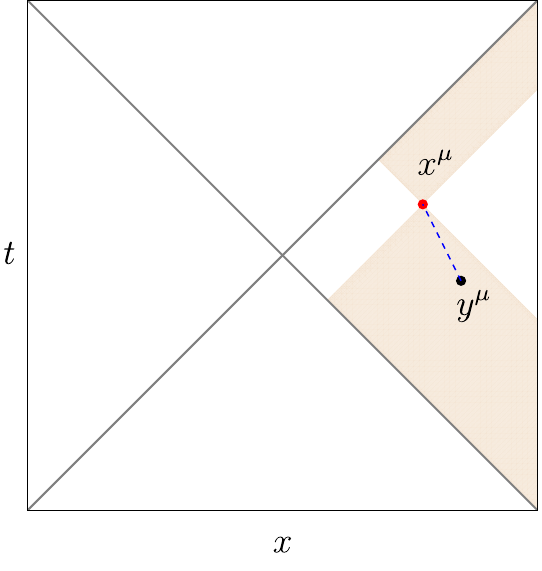}
    \caption{}\label{gpxy1}
  \end{subfigure}
  \hspace{0.5cm}
  \begin{subfigure}{0.3\textwidth}
    \includegraphics[width=\linewidth]{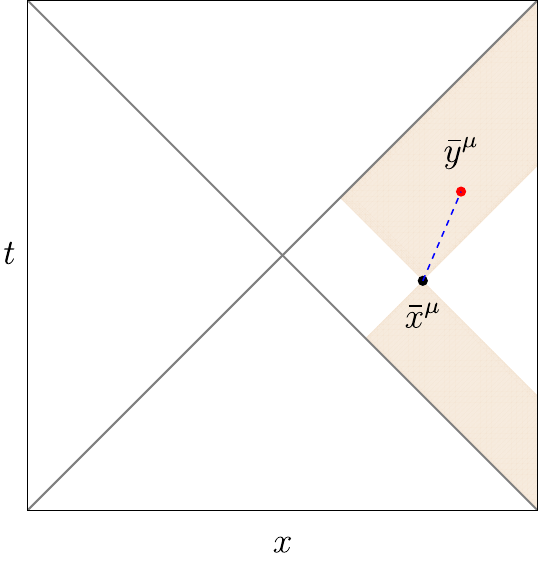}
    \caption{}\label{gpxy2}
  \end{subfigure}
  \hspace{0.5cm}
  \begin{subfigure}{0.3\textwidth}
    \includegraphics[width=\linewidth]{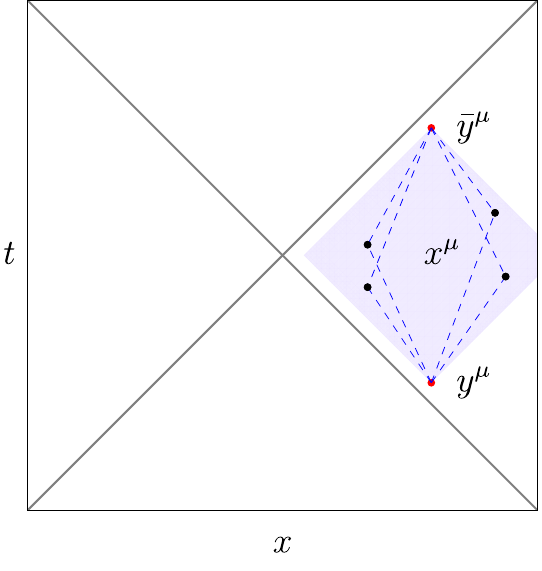}
    \caption{}\label{kvrsse}
  \end{subfigure}
  \caption{(a) Path 1: From $y^\mu\in R$ to $x^\mu\in R$. (b) Path 2: From $\bar{x}^\mu\in R$ to $\bar{y}^\mu\in R$.\hspace{2cm} (c) Paths 1 and 2, where $x^\mu = \bar{x}^\mu$ and $y^\mu$ is related to $\bar{y}^\mu$ by a time reflection.}
\end{figure}

 Let us consider the configurations shown in Figs.~\ref{gpxy1} and \ref{gpxy2}, where Path~1 corresponds to the propagation from the event $y^\mu \equiv (y^0, y^1)$  to $x^\mu \equiv (x^0, x^1)$, while Path~2 corresponds to the propagation from $\bar{x}^\mu \equiv (\bar{x}^0, \bar{x}^1)$ to $\bar{y}^\mu \equiv (\bar{y}^0, \bar{y}^1)$. Notice that both propagations take place within the right Rindler wedge $R$.To describe the Feynman propagator from the perspective of a Rindler observer, we employ the coordinate transformation between Minkowski coordinates—restricted to the right wedge—and Rindler coordinates $(\eta, \rho)$, given by the standard form of Eq.~\eqref{trrhoeta1}, that is, with the identifications $2/\alpha = a$ and $\tilde{\alpha} a = 1$.

Thus, we obtain the following squared spacetime intervals:
\begin{eqnarray}
\text{Path 1:}\quad\sigma^2(x,y) &=& -(x^{0} - y^{0})^2 + (x^{1} - y^{1})^2 \nonumber\\
&=& \rho_{x}^2 + \rho_{y}^2 - 2\rho_{x} \rho_{y} \cosh\qty[a(\eta_{x} - \eta_{y})],\label{ietm12} \\
\text{Path 2:}\quad\sigma^2(\bar{x},\bar{y}) &=& -(\bar{y}^{0} - \bar{x}^{0})^2 + (\bar{y}^{1} - \bar{x}^{1})^2 \nonumber\\
&=& \bar{\rho}_{x}^2 + \bar{\rho}_{y}^2 - 2\bar{\rho}_{x} \bar{\rho}_{y} \cosh\qty[a(\bar{\eta}_{x} - \bar{\eta}_{y})].\label{ietm122}
\end{eqnarray}

As expected, the temporal dependence arises solely through the intervals $\Delta\eta = \eta_x - \eta_y$ and $\Delta\bar{\eta} = \bar{\eta}_x - \bar{\eta}_y$, reflecting the static nature of the Rindler geometry. To compare the two propagators, we consider the setup $x^\mu = \bar{x}^\mu$ and $\bar{y}^\mu = (-y^0, y^1)$, as shown in Fig.~\ref{kvrsse}, which implies $\bar{\eta}_x = \eta_x$, $\bar{\rho}_x = \rho_x$, $\bar{\eta}_y = -\eta_y$, and $\bar{\rho}_y = \rho_y$. Consequently, $\Delta\eta = \eta_x - \eta_y$ and $\Delta\bar{\eta} = \eta_x + \eta_y$. Substituting these relations into Eqs.~\eqref{ietm12} and~\eqref{ietm122}, we obtain:
\begin{eqnarray}
\text{Path 1:}\quad\sigma^2(x,y) &=& \rho_{x}^2 + \rho_{y}^2 - 2\rho_{x} \rho_{y} \cosh\qty[a(\eta_{x} - \eta_{y})],\label{rinc2inr1}\\
\text{Path 2:}\quad\sigma^2(x,\bar{y}) &=& \rho_{x}^2 + \rho_{y}^2 - 2\rho_{x} \rho_{y} \cosh\qty[a(\eta_{x} + \eta_{y})].\label{rinc2inr2}
\end{eqnarray}

The frequency-space propagator associated with Path 1 is then:
\begin{eqnarray}
\tilde{G}_F^{\text{Path 1}}(\Omega, \rho_{x}, \rho_{y}) &&= \int_{-\infty}^{\infty} d(\Delta \eta)\, G_F^{\text{Path 1}}(\Delta\eta, \rho_{x}, \rho_{y})\, e^{-i\Omega \Delta\eta}, \nonumber\\
&&= \int_0^\infty \frac{ds}{4\pi s} \exp\qty[\frac{i}{4s}(\rho_{x}^2 + \rho_{y}^2) - i s m^2 - s \epsilon] \qty[\frac{2}{a} K_{i\Omega/a}\qty(\frac{i \rho_{x} \rho_{y}}{2s})]. \label{ataumnor}
\end{eqnarray}

Here we have used Eq \eqref{gradeqeq} with the identifications $\zeta = i \rho_{x} \rho_{y} / (2s)$, $\gamma = e^{a \Delta\eta}$, $z = 1$, and $\nu = i \Omega / a$.\\

Similarly, for Path 2 we have:
\begin{eqnarray}
    \tilde{G}_F^{\text{Path 2}}(\bar{\Omega}, \rho_{x}, \rho_{y}) &&= \int_{-\infty}^{\infty} d(\Delta \bar{\eta})\, G_F^{\text{Path 2}}(\Delta\bar{\eta}, \rho_{x}, \rho_{y})\, e^{-i\bar{\Omega} \Delta\bar{\eta}}.
\end{eqnarray}
From Eqs.~\eqref{rinc2inr1} and~\eqref{rinc2inr2}, and noting that $\Delta\eta$ and $\Delta\bar{\eta}$ are dummy integration variables, it follows that setting $\bar{\Omega} = \Omega$ makes the two expressions identical. Thus, we find
\begin{equation}
 \tilde{G}_F^{\text{Path 1}}(\Omega, \rho_{x}, \rho_{y}) 
 = \tilde{G}_F^{\text{Path 2}}(\Omega, \rho_{x}, \rho_{y}).\label{case1path12R}
\end{equation}

This equality shows that propagation between events within the Rindler wedge $R$, where the time evolution generated by $\partial_\eta$ and the corresponding interval $\Delta\eta$ are well defined, is physically equivalent for Paths~1 and~2. This result is fully consistent with the viewpoint of an inertial observer in Minkowski spacetime.

\subsubsection{Propagation from the wedge $P$ to $R$, and from $R$ to $F$}

\begin{figure}[h]
 \centering
  \begin{subfigure}{0.3\textwidth}
    \includegraphics[width=\linewidth]{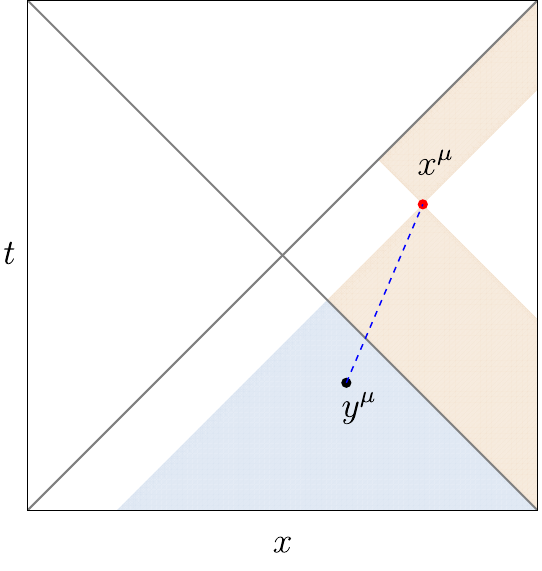}
    \caption{}\label{gpxy11}
  \end{subfigure}
  \hspace{0.5cm}
  \begin{subfigure}{0.3\textwidth}
    \includegraphics[width=\linewidth]{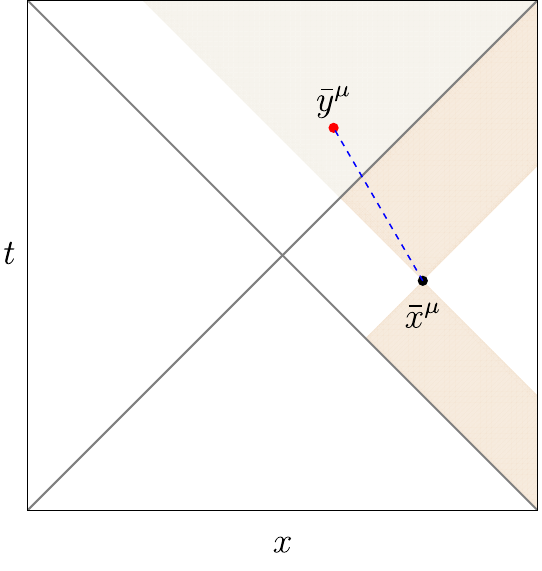}
    \caption{}\label{gpxy22}
  \end{subfigure}
  \hspace{0.5cm}
  \begin{subfigure}{0.3\textwidth}
    \includegraphics[width=\linewidth]{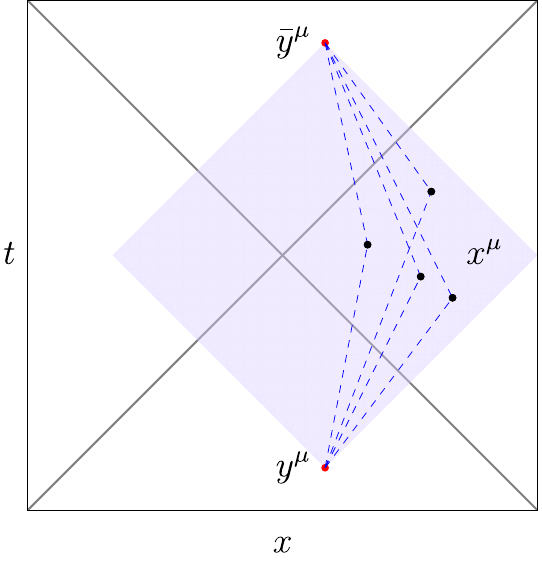}
    \caption{}\label{kvrssecd}
  \end{subfigure}
  \caption{(a) Path 1: From $y^\mu \in P$ to $x^\mu \in R$. (b) Path 2: From $\bar{x}^\mu \in R$ to $\bar{y}^\mu \in F$.\hspace{2cm} (c) Paths 1 and 2, where $x^\mu = \bar{x}^\mu$ and $y^\mu$ is related to $\bar{y}^\mu$ by a time reflection.}
\end{figure}

Let us consider the configurations shown in Figs.~\ref{gpxy11} and~\ref{gpxy22}, where Path~1 represents propagation from the event $y^\mu \equiv (y^0, y^1) \in P$ to $x^\mu \equiv (x^0, x^1) \in R$, and Path~2 corresponds to propagation from $\bar{x}^\mu \equiv (\bar{x}^0, \bar{x}^1) \in R$ to $\bar{y}^\mu \equiv (\bar{y}^0, \bar{y}^1) \in F$. The left wedge $L$ is excluded from this analysis, as there is no causal connection—according to an inertial observer in Minkowski spacetime—between points located in wedges $R$ and $L$. To carry out this analysis, we employ the Rindler coordinate charts given by the standard forms ($2/\alpha = a$ and $\tilde{\alpha} a = 1$) of Eqs.~\eqref{trrhoeta1} and~\eqref{trrhoeta3}–\eqref{trrhoeta4}.\\

Thus, we obtain the following squared spacetime intervals:
\begin{eqnarray}
\text{Path 1:}\quad\sigma^2(x,y)&=& -(x^0 - y^0)^2 + (x^1 - y^1)^2\nonumber\\
&=& -\rho_p^2 + \rho_r^2 - 2\rho_p\rho_r \sinh\qty[a(\eta_r - \eta_p)],\label{screw1}\\
\text{Path 2:}\quad\sigma^2(\bar{x}, \bar{y})&=& -(\bar{y}^0 - \bar{x}^0)^2 + (\bar{y}^1 - \bar{x}^1)^2\nonumber\\
&=&  -\bar{\rho}_f^2 + \bar{\rho}_r^2 + 2\bar{\rho}_f \bar{\rho}_r \sinh\qty[a(\bar{\eta}_r - \bar{\eta}_f)].\label{screw2}
\end{eqnarray}

To compare both propagators, we identify $\bar{x}^\mu = x^\mu$ and $\bar{y}^\mu \equiv (-y^0, y^1)$, as illustrated in Fig.~\ref{kvrssecd}. Consequently, we have $\bar{\eta}_r = \eta_r$, $\quad \bar{\rho}_r = \rho_r$, $\quad \bar{\eta}_f = -\eta_p$, and $\quad \bar{\rho}_f = \rho_p$, so that $\Delta\eta = \eta_r - \eta_p$ and $\Delta\bar{\eta} = \eta_r + \eta_p$. Thus, the squared spacetime intervals in Eqs.~\eqref{screw1} and \eqref{screw2} then become:
\begin{eqnarray}
\text{Path 1:}\quad\sigma^2(x,y) &=& -\rho_p^2 + \rho_r^2 - 2\rho_p\rho_r \sinh\qty[a(\eta_r - \eta_p)],\\
\text{Path 2:}\quad\sigma^2(x,\bar{y}) &=& -\rho_p^2 + \rho_r^2 + 2\rho_p\rho_r \sinh\qty[a(\eta_r + \eta_p)].
\end{eqnarray}

Accordingly, the frequency-space propagator for Path 1 is given by

\begin{eqnarray}
\tilde{G}_F^{\text{Path 1}}(\Omega, \rho_p, \rho_r) &&= \int_{-\infty}^{\infty} d\eta_r\, G_F^{\text{Path 1}}(\eta_r-\eta_p,\rho_p, \rho_r)e^{-i\Omega\eta_r}\nonumber\\
&&= \int_{-\infty}^{\infty} d(\Delta \eta)\, G_F^{\text{Path 1}}(\Delta\eta,\rho_p, \rho_r)e^{-i\Omega \Delta\eta}e^{-i\Omega\eta_p}\nonumber\\
&&\hspace{-3cm}= \int_0^\infty \frac{ds}{4\pi s} \exp\qty[\frac{i}{4s}(-\rho_p^2 + \rho_r^2) - i s m^2 - s \epsilon]\qty[\frac{2e^{-i\Omega\eta_p}e^{-\frac{\pi\Omega}{2a}}}{a}K_{-i\Omega/a}\qty(-\frac{\rho_p\rho_r}{2s})],\label{ataumnor}
\end{eqnarray}
where we used the identity given in Eq. \eqref{formulmaster} with the substitutions $\zeta = -\rho_p \rho_r / (4s)$, $\gamma = e^{a\Delta\eta}$, and $\nu = -\Omega/a$.\\

As seen from Eq.~\eqref{ataumnor}, and noting that the Macdonald function $K_{i\nu}(2\zeta)$ is even in its order, we obtain the identity:
\begin{equation}
    \tilde{G}_F^{\text{Path 1}}(-\Omega, \rho_p, \rho_r) = e^{\pi\Omega/a}\, \tilde{G}_F^{\text{Path 1}}(\Omega, \rho_p, \rho_r).\label{bolztfa}
\end{equation}

Similarly, for Path 2 we obtain:
\begin{eqnarray}
\tilde{G}_F^{\text{Path 2}}(\bar{\Omega}, \rho_p, \rho_r) &=& \int_{-\infty}^{\infty} d\eta_r\, G_F^{\text{Path 2}}(\eta_r+\eta_p,\rho_p, \rho_r)\, e^{-i\bar{\Omega} \eta_r}\nonumber\\
&=& \int_{-\infty}^{\infty} d(\Delta \bar{\eta})\, G_F^{\text{Path 2}}(\Delta\bar{\eta},\rho_p, \rho_r)\, e^{-i\bar{\Omega} \Delta\bar{\eta}}e^{i\bar{\Omega}\eta_p}\nonumber\\
&&\hspace{-3cm}= \int_0^\infty \frac{ds}{4\pi s} \exp\qty[\frac{i}{4s}(-\rho_p^2 + \rho_r^2) - i s m^2 - s \epsilon]\qty[\frac{2e^{i\bar{\Omega}\eta_p}e^{\frac{\pi\bar{\Omega}}{2a}}}{a}K_{i\bar{\Omega}/a}\qty(-\frac{\rho_p\rho_r}{2s})],
\end{eqnarray}
where we have applied Eq.~\eqref{formulmaster} with $\zeta = -\rho_p \rho_r / (4s)$, $\gamma = e^{a\Delta\bar{\eta}}$, and $\nu = \Omega/a$. Setting $\bar{\Omega} = \Omega$, we finally obtain:
\begin{equation}
     \tilde{G}_F^{\text{Path 2}}(\Omega, \rho_p, \rho_r)= \tilde{G}_F^{\text{Path 1}}(-\Omega, \rho_p, \rho_r).
\end{equation}

Therefore, from Eq.~\eqref{bolztfa}, the squared modulus of the ratio between the two frequency-space propagators is
\begin{equation}
    \abs{\frac{\tilde{G}_F^{\text{Path 1}}(\Omega, \rho_p, \rho_r)}{\tilde{G}_F^{\text{Path 2}}(\Omega, \rho_p, \rho_r)}}^2 = e^{-\beta\Omega}\quad,\quad\beta=2\pi/a.\label{path12eabse}
\end{equation}

This ratio follows a Boltzmann factor with an effective temperature equal to the Unruh value, $T = a/(2\pi)$. From the viewpoint of the observer confined to wedge $R$, Path~1—propagation from $P$ into $R$—corresponds to the emission of scalar quanta from the past horizon, while Path~2—propagation from $R$ into $F$—corresponds to their absorption at the future horizon. Thus, although propagations across causal horizons are physically equivalent for an inertial observer, they appear to the Rindler observer as a clear manifestation of thermality associated with the horizons—an effect otherwise hidden within the inertial propagator.

\section{Discussion}
In this article, we introduced a method in which (1) the Feynman propagator was employed to detect the inequivalence of physical processes across causal horizons from the perspective of observers constrained to regions that are causally disconnected from the rest of Minkowski spacetime; (2) the Fourier transform was implemented following the convention used in the analysis of vacuum fluctuations; and (3) a specific setup was employed involving the triad of events $x^\mu$, $y^\mu$, and $\bar{y}^\mu$, with $y^0 < x^0 < \bar{y}^0$. Moreover, $y^\mu$ and $\bar{y}^\mu$ are related by a reflection in the Minkowski time coordinate. This configuration enabled us to compute the ratio of propagators across the past and future horizons in the frequency domain.

Following this method, we first analyzed in detail the case of the causal diamond region $D_R$. By construction, the procedure imposes no restrictions on the relative distance of events from the causal horizon, allowing us—without loss of generality—to adopt the near-horizon approximation where the effective geometry becomes static. This simplification enabled a consistent definition of the Fourier transform in the interval $\Delta\eta$ and revealed that propagation across the causal horizons can be causally interpreted as the emission and absorption of scalar quanta at the past and future horizons, respectively. The ratio between these two processes yields a Boltzmann factor with a temperature precisely equal to that associated with the causal diamond. These results show that thermality in quantum field theory can arise solely from causal structure, indicating that causal diamonds—as minimal units of spacetime causality—inherently encode thermodynamic behavior independent of acceleration or gravitation.

Finally, for completeness, we extended the analysis to the right Rindler wedge $R$. Using the same causal setup between events in the regions $P$, $R$, and $F$, we obtained an analogous interpretation: the ratio between the emission and absorption processes reproduces the Boltzmann factor with the familiar Unruh temperature, $T = a/(2\pi)$, as expected.

\begin{acknowledgments} We would like to thank the attendees of the 2025 TAMU-Princeton-Caspar-Baylor-CSU-UIUC Summer School, where these and other results were first presented, for their interest and many questions. We are especially grateful to Wolfgang Schleich for his careful reading of a draft version of the paper, for his thoughtful questions and suggestions and for his encouragement to publish these results. All the authors of this paper were partially supported by the Army Research Office (ARO) under Grant No. W911NF-23-1-0202. In addition, G.V.-M. gratefully acknowledges the Center for Mexican American and Latino/a Studies at the University of Houston for their generous support through a Lydia Mendoza Fellowship. \end{acknowledgments}

\bibliography{apssamp}

\end{document}